# Recovering Phosphine in Venus' Atmosphere from SOFIA Observations

## Comment on

### "*Phosphine in the Venusian Atmosphere: A Strict Upper Limit from SOFIA GREAT Observations*"

**Jane S. Greaves[1], Janusz J. Petkowski[2,3], Anita M. S. Richards[4], Clara Sousa-Silva[5], Sara Seager[2,6,7] and David L. Clements[8]**

[1] CHART, School of Physics & Astronomy, Cardiff University, 4 The Parade, Cardiff CF24 3AA, UK.

[2] Department of Earth, Atmospheric and Planetary Sciences, Massachusetts Institute of Technology, 77 Massachusetts Avenue, Cambridge, MA 02139, USA.

[3] JJ Scientific, 02-792 Warsaw, Poland.

[4] Jodrell Bank Centre for Astrophysics, Department of Physics and Astronomy, The University of Manchester, Manchester, UK.

[5] Bard College, Campus Road, Annandale-on-Hudson, NY 12504, USA.

[6] Department of Physics, Massachusetts Institute of Technology, 77 Massachusetts Avenue, Cambridge, MA 02139, USA.

[7] Department of Aeronautics and Astronautics, Massachusetts Institute of Technology, 77 Massachusetts Avenue, Cambridge, MA 02139, USA.

[8] Department of Physics, Imperial College London, London, UK.

Corresponding authors: Jane S. Greaves (greavesj1@cardiff.ac.uk), Janusz J. Petkowski (jjpetkow@mit.edu).

**Key Points:**

- We recover Venusian phosphine in *SOFIA* spectra by reducing contaminating signals; the $PH_3$ abundance is ~3 part-per billion (ppb).

- Six recoveries/limits show $PH_3$ depleting between clouds and mesosphere, which would require an unknown re-formation process or extra source.

- Recoveries and upper limits can instead be reconciled by $PH_3$ photolysis, as high/low abundances correspond to Venusian mornings/evenings.

**Abstract**

*Searches for phosphine in Venus' atmosphere have sparked a debate. Cordiner et al. 2022 analyse spectra from the Stratospheric Observatory For Infrared Astronomy (SOFIA) and infer <0.8 ppb of PH₃. We noticed that some spectral artefacts arose from non-essential calibration-load signals. By-passing these signals allows simpler post-processing and a 5.7σ candidate detection, suggesting ~3 ppb of PH₃ above the clouds. Compiling six phosphine results hints at an inverted abundance trend: decreasing above the clouds but rising again in the mesosphere from some unexplained source. However, no such extra source is needed if phosphine is undergoing destruction by sunlight (photolysis), to a similar degree as on Earth. Low phosphine values/limits are found where the viewed part of the super-rotating Venusian atmosphere had passed through sunlight, while high values are from views moving into sunlight. We suggest Venusian phosphine is indeed present, and so merits further work on models of its origins.*

**Plain Language Summary**

*Cordiner et al. find no phosphine in Venus' atmosphere, using the airborne SOFIA telescope. By-passing some instrumental effects, we extract a detection with 5.7σ-confidence from the same data. We can resolve the tension between high and low PH₃ abundance values by noticing that the former are from 'mornings' in Venus' atmosphere and the latter from 'evenings'. Sunlight reduces the amount of phosphine in Earth's atmosphere by an order of magnitude, so similarly on Venus, we might expect lower abundances in data taken when the part of the atmosphere observed has passed through sunlight. If the six available datasets can be reconciled in this way, further modelling of possible sources of PH₃ (e.g. volcanic, disequilbrium chemistry, extant life) seem worthwhile.*

**1 Introduction**

Phosphine, if present in Venus' atmosphere, would be unexpected on an oxidised planet. Greaves et al. (2021) searched for $PH_3$ absorption at 1 mm wavelength, testing the concept that this molecule may be a biosignature when seen in anoxic environments. The unexpected detection-candidates from *JCMT* and *ALMA* have stimulated much community work on robust spectral processing, and on other methods to detect $PH_3$ at Venus, mostly proving negative except for an in-situ mass-spectrometry recovery (Mogul, Limaye, Way, et al., 2021). Particularly deep (above-cloud) limits have been set by infrared spectroscopy (Encrenaz et al., 2020; Trompet et al., 2020).

We comment here on the findings of Cordiner et al. (2022), hereafter C22, who present a deep upper limit from $PH_3$ observations with the *GREAT* instrument on *SOFIA*. They propose that *all* the candidate detections of phosphine in Venus' atmosphere could in fact be null results, given the complexity of the challenging observations – although their Figure 1 omits findings by Greaves et al. (2022), our work after calibration and contamination issues were fully resolved, where we find self-consistent 6-8σ detections for the $PH_3$ J=1-0 line from *JCMT* and *ALMA*.

The C22 observations are of the rotational transitions J=4-3 and 2-1 (around 1 and 0.5 THz), uniquely accessible to the *SOFIA* airborne telescope, and complementary to the existing J=1-0 spectra (at 0.27 THz). From their J=4-3 data processing, C22 find an upper limit of 0.8 ppb of $PH_3$, applicable to most of the planet and 75+ km altitudes, while their J=2-1 results suggest ~2.3 ppb could be present but only with 1.5σ confidence. These abundances are difficult to reconcile with ~20 ppb levels from the J=1-0 data, without invoking strong temporal-variations or steep gradients over the slightly different altitudes these lines trace.

## 2 Materials and Methods

C22 note the existence in the *GREAT* spectra of quasi-periodic fringe patterns, due to standing waves between optical elements and to frequency-dependent gain factors used in calibration. Their calibration to antenna temperatures $T_A$ follows the standard method of dividing the power difference of on- and off-Venus spectra by the power difference of hot and cold calibration-load signals, and then multiplying by the temperature difference of the hot and cold loads. We noticed that much of the fringing is introduced because the standing waves differ when observing the sky and the calibration loads. However, calibration to $T_A$ is not essential in measuring the line-to-continuum ratios, $l/c$, from which abundances derive. In the case of the PH$_3$ J=4-3 line components (seen by the "4G2 pixel"), an alternative is

$$l/c = (On_{line} - Off^*_{line}) / [0.5(On - Off)] \qquad [1]$$

where *On* and *Off* are the spectra on Venus and on adjacent blank sky, the subscript *line* indicates the broadband signals have been subtracted, and *Off$_{line}$*\* represents the instrumental line-signal that *GREAT* would see for a featureless patch of sky of similar brightness to Venus. *Off$_{line}$*\* was generated by multiplying *Off$_{line}$* by a factor ~1.05 and adjusting this scalar until the residual *(On$_{line}$ -Off$_{line}$\*)* was minimised – that is, the procedure tests the null hypothesis, that no absorbing gases are present. Smooth fits to *On* and *Off* (the continuum signals) were used in the denominator of [1] to further minimise noise. The factor of 0.5 arises because *GREAT* is a double-sideband instrument with approximately equal sideband gains (see C22), and so records the planetary continuum twice. The Eq. [1] method has worked well here for PH$_3$ J=4-3, reaching a similar noise level to C22, but our approach failed for the PH$_3$ J=2-1 line-pair ("4G1 pixel") because the ripples differ between *On$_{line}$* and *Off$_{line}$*.

Remaining ripples in the J=4-3 spectra were then removed by a one-stage Fourier process, contrasting to the iterative 7-step Lomb-Scargle periodogram approach used by C22 (or traditional polynomial fitting, which is less useful for spectra with many ripples). Both we and C22 similarly "masked" the spectral regions where the four PH$_3$ components lie, to avoid fitting real lines as if they are ripples. C22's periodogram method works intrinsically on masked data, while we interpolated across the line regions with quadratic fits anchored on adjacent spectral pixels. We used 3-sigma cuts in Fourier space, with features above these cuts inverse-Fourier-transformed to create a family of model sinusoids. Subtracting these model baselines yields well-flattened spectra (Figures 1a, 1b). Finally we "stack" the 24 samples of PH$_3$ absorption in the data (Figure 1c), namely from the 6 observations and their 4 spectral sections covering the J=4-3 components, as these are of similar intrinsic line-depth (see Figure 3 in C22). Two of the PH$_3$ features are close together (Figure 1b), and so to avoid duplications, we replaced the secondary occurrences with noise values from elsewhere in the spectrum, before making the final stack.

Several robustness checks were run, exploring possible processing issues.

- The net result could be dominated by a few strong artefacts. This was found not to be the case, with the line-integrals from the 24 samples following an approximately normal distribution. The result is also robust to removing a few points. For example, the final observation (#040402) was the noisiest, and removing all these samples from the stack shifts the net line-integral by only -20%.

- There could be terrestrial atmospheric signals in the band, that affect the validity of minimising the numerator in Eq. [1]; in particular, C22 note two frequencies where there may be $O_3$ absorption. We re-calculated the scalar used in Eq. [1] with the $O_3$ regions blanked, and found negligible change (around 1% reduction in noise).
- The interpolation across the $PH_3$ line-regions might not be following the correct trend, potentially always producing negative residuals that mimic absorption lines. We tested this by re-running the processing identically, but shifting the spectral sections to parts of the band without phosphine features. From 50 tests, none produced a result like Figure 1c, namely a "fake" absorption that is the only feature, centrally placed, and of comparable width to line models (see below).

## 3 Results

Three of the four expected $PH_3$ J=4-3 components are visible when all the observations are co-added (Figure 1b), while only one component was apparent in Figure 3 of C22. In Figure 1c, our final stacked spectrum indicates an overall $5.7\sigma$ detection of $PH_3$ J=4-3, when integrated over $\pm17$ MHz (the masked region). This confidence level changes marginally (by $\pm0.4\sigma$) if a different range of spectral pixels is used to calculate the zero-level.

We then modelled our net spectrum using the same online tool as C22. We ran a model for 1 ppb of phosphine and scaled it linearly for different abundances, and then calculated a reduced-chi-squared statistic to assess goodness of fit. Figure 1c illustrates that 3 ppb of $PH_3$ provides a good match to the observed line, with $\chi^2_r$ of 0.75. (The uncertainties used in $\chi^2_r$ were generated per spectral pixel from the internal data dispersion.) We consider that fitting the stacked spectrum better mitigates against artefacts, compared to matching the individual components (Figure 1b) against the model. For example, inserting a single positive "spike" near the highest-frequency component (Figure 1b) was found to significantly reduce the inferred $PH_3$ abundance, as the $\chi^2_r$ test attempts to minimise the discrepancy of positive data against a negative model-line. As a positive feature does appears here in this part of C22's spectrum (their Figure 3), this could have driven their upper limit down to the $\leq$0.8 ppb they obtain. However, we do not rule out an abundance as low as 0.8 ppb, which is at the lower 99% confidence-bound in our $\chi^2_r$ tests. We also note that C22 estimated ~2.3 ppb from their $PH_3$ J=2-1 spectrum, albeit at only $+1.5\sigma$ confidence, and this estimate is compatible with our 3 ppb result.

C22 (Figure S4) find that altitudes around ~80 km are the best-sampled at the $PH_3$ J=4-3 line-frequency; our recovery here is consistent with the model that we both use, and with the predicted short lifetime of $PH_3$ above ~80 km (Bains, Petkowski, Seager, et al., 2021). Altitudes are however uncertain because the $PH_3$-$CO_2$ pressure-broadening coefficient has not been experimentally verified. We also note that all GHz/THz data are limited by the spectral span that can be recovered. Here, any absorption wider than ~200 MHz leads to merged $PH_3$ J=4-3 components, and so any phosphine signatures below ~70 km (roughly cloud-top level) are lost.

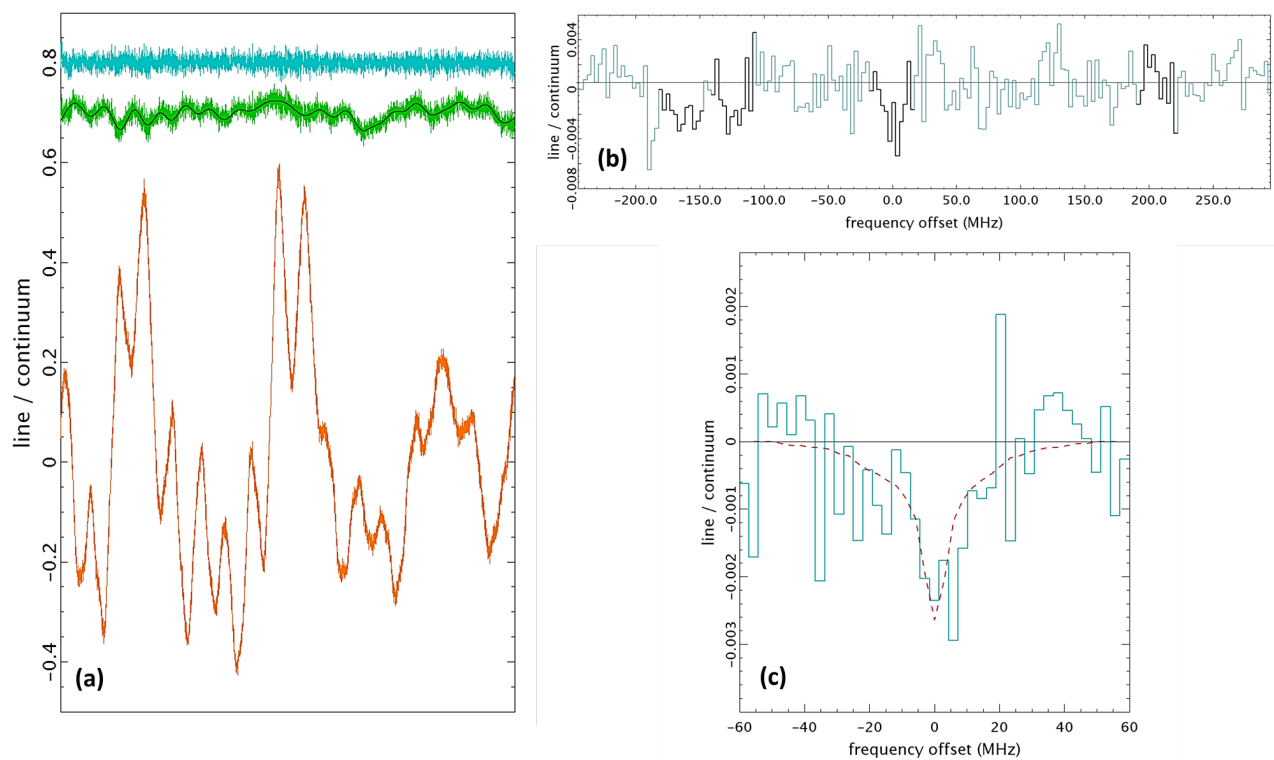

Figure 1. *The process of extracting the PH$_3$ J=4-3 signal from the GREAT data is illustrated. Panel (a) illustrates processing of 6000 spectral pixels from observation #040093 from the first SOFIA flight (the reference spectral pixel is mid-band; vertical offsets are for clarity only). The lower orange histogram is from a standard On-Off processing; the middle green histogram is the result after applying the Eq.[1] step, and is overlaid with the Fourier-derived trend (black curve); the top blue histogram is the flattened output after subtracting this trend. In panel (b), the blue histogram is the unweighted average of all six observations made over the three flights, with the sections containing the four PH$_3$ J=4-3 components highlighted in black. The spectral pixels are binned in groups of 12 in (b) and (c) to improve clarity, and the thin black lines show the zero-level correction made between (b) and (c). Panel (c) shows (blue histogram) the unweighted stack of all 24 spectral sections containing PH$_3$ J=4-3 features. The dashed red curve is a model for 3 ppb of phosphine, generated vis the Planetary Spectrum Generator (PSG) (https://psg.gsfc.nasa.gov/index.php) following C22. The four line-components from the PSG model were masked and stacked similarly to the data.*

## 4 Discussion

Debates continue about the best methods to acquire and process deep GHz/THz spectra of Venus. These observations are very challenging in dynamic range, as Venus is so bright, revealing "ripples" in spectral baselines that are not evident in more typical telescope usage. Depending on preferred approaches, different authors argue for between zero and three published detections of rotational (J) transitions of PH$_3$.

We can compare results from the data discussed here with the outcomes of other searches for phosphine at Venus, and assess whether this results in a plausible altitude profile of the molecule (Figure 2). The trend found by connecting the results from six searches for phosphine appears as an upwards decline that then reverses, i.e. PH$_3$ that is depleted somewhere between ~50 km and ~80 km. This is hard to explain in the absence of a chemical route to reform the molecules, or a new mesospheric source. The order-of-magnitude contrast between some of the candidate detections and the upper limits has led to doubts over the presence of phosphine.

However, we noticed that this divide is also between observations made when the 'morning' versus the 'evening' sides of Venus' atmosphere were targeted – and this is relevant in gas-mixing processes (e.g. (Lefèvre et al., 2022)). Where the gas observed on Venus has travelled through sunlight and is descending towards the night-side of the planet, we detect at most the ~3 ppb of phosphine estimated here. In contrast, where gas is rising into sunlight, we observe ≥ ~20 ppb of PH₃. Hence photolysis – similar to the observed destruction of terrestrial phosphine by sunlight (Sousa-Silva et al., 2020) – could explain the split between high and low phosphine abundances observed on Venus. It is striking that this evening/morning difference is comparable in magnitude to the factor ~10 difference over night/day for phosphine in the Earth's atmosphere (Glindemann et al. 1996; noting absolute abundances are much lower on Earth).

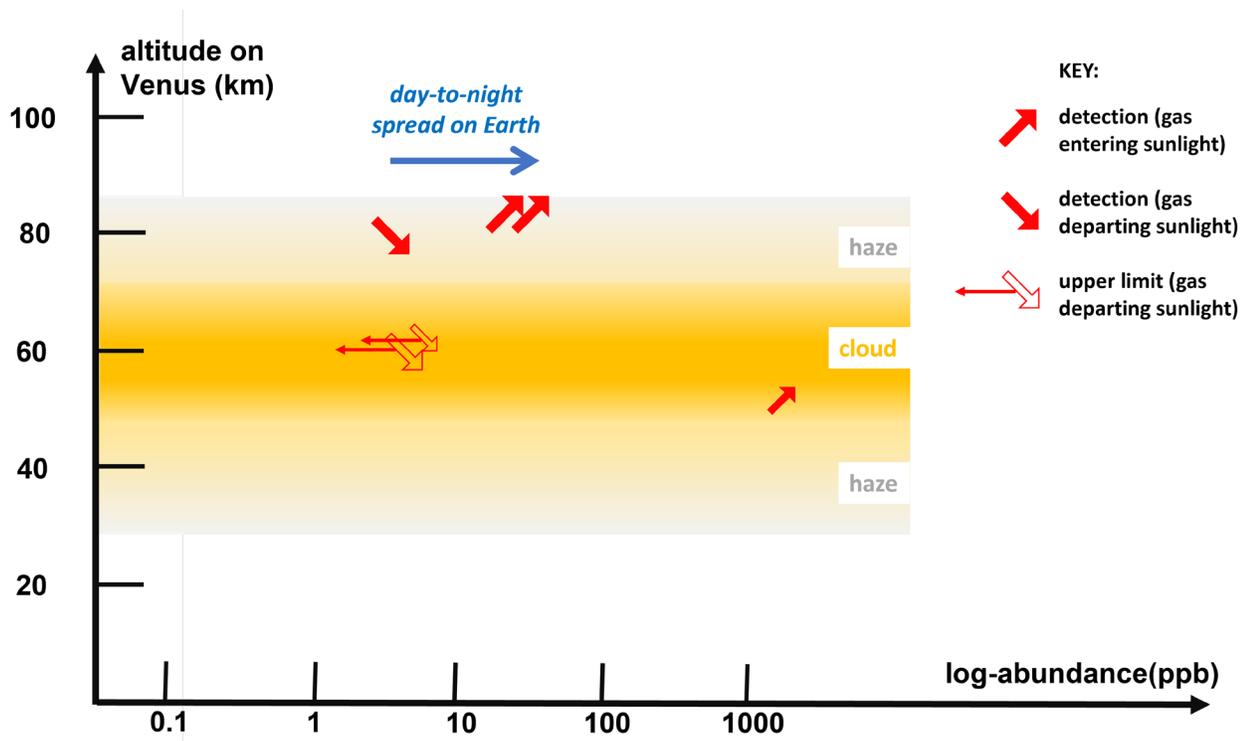

*Figure 2. The trend of phosphine abundances by altitude is sketched. Symbols indicate candidate detections plus best upper limits for phosphine abundances. Rising arrows indicate observations made where the super-rotating atmosphere was rising into sunlight and falling arrows indicate observations made where the atmosphere was descending towards the nightside (see key). Large and small symbols indicate that a large fraction of the planet area was observed, or that a small region was sampled, respectively. Abundance estimates are, from top: ~20, 25 ppb from J=1-0 data (via* (Greaves et al., 2022) *and with altitude proposed by C22); 3 ppb from J=4-3 data (this work; beam centred on the evening side); <7 ppb at 62 km from 4 μm spectra (Trompet et al., 2020: low-latitude data to best match whole-planet studies); <5 ppb at 60 km from 10 μm spectra (Encrenaz et al., 2020: latitudes within ±50°); ~2 ppm at 51 km from Pioneer-Venus in-situ sampling during descent* (Mogul, Limaye, Way, et al., 2021). *The blue arrow indicates the ten-fold increase of terrestrial phosphine from day to night (Glindemann et al. 1996) – note the arrow's plotted position is arbitrary; Earth hosts much lower PH₃ than Venus.*

## 5 Conclusions

The question regarding phosphine in Venus' atmosphere is likely to be debated for some time. A further *JCMT* survey[1] is ongoing, producing open-source data that should yield more definitive answers – in particular, that team is now processing broadband spectra that can sample the cloud decks. The most direct answer regarding phosphine could come from new in-situ sampling, potentially with the addition of one laser channel to the Venus Tunable Laser Spectrometer (VTLS) instrument on-board the *DAVINCI* descent probe (Garvin et al. 2022).

The origins of any phosphine present are also debated, and most scenarios are hard to test for lack of some contextual data. For example, it seems only extraordinary volcanic activity could make ~ppb-level phosphine (Bains et al., 2022) but vulcanism on Venus is not well understood. In some new avenues, (Ferus et al., 2022) discuss abiotic routes to phosphine involving redox disequlibrium, while others (Bains, Petkowski, Rimmer, et al., 2021; Mogul, Limaye, Lee, et al., 2021) explore phototrophic life and the habitability of the clouds. We conclude that establishing an improved $PH_3$ altitude-profile is worthwhile to test these new models of origins.

## Acknowledgments

This work is based in part on observations made with the NASA/DLR *Stratospheric Observatory for Infrared Astronomy* (*SOFIA*). *SOFIA* is jointly operated by the Universities Space Research Association, Inc. (USRA), under NASA contract NNA17BF53C, and the Deutsches SOFIA Institut (DSI) under DLR contract 50 OK 2002 to the University of Stuttgart. We thank Helmut Wiesemeyer, Martin Cordiner and staff at the *SOFIA* Science Center for their invaluable help.

## Open Research

The SOFIA Level 1 data are available under project id 75_0059_1 through the public data archive at https://irsa.ipac.caltech.edu/applications/sofia. The custom software to generate the data shown in the figures is supplied at https://zenodo.org/record/7692288#.ZAC363bP3IU. The script requires the UK-Starlink software (Currie et al., 2014) which is currently supported by the East Asian Observatory and available at https://starlink.eao.hawaii.edu/starlink/2021ADownload.

## References

Bains, W., Petkowski, J. J., Seager, S., Ranjan, S., Sousa-Silva, C., Rimmer, P. B., et al. (2021). Phosphine on Venus Cannot be Explained by Conventional Processes. *Astrobiology*, *21*(10), 1277–1304. Retrieved from https://ui.adsabs.harvard.edu/abs/2020arXiv200906499B

Bains, W., Petkowski, J. J., Rimmer, P. B., & Seager, S. (2021). Production of Ammonia Makes Venusian Clouds Habitable and Explains Observed Cloud-Level Chemical Anomalies. *Proceedings of the National Academy of Science*, *118*(52).

Bains, W., Shorttle, O., Ranjan, S., Rimmer, P. B., Petkowski, J. J., Greaves, J. S., & Seager, S. (2022). Constraints on the production of phosphine by Venusian volcanoes. *Universe*, *8*(1), 54.

---

[1] https://www.eaobservatory.org/jcmt/science/large-programs/jcmt-venus-monitoring-phosphine-and-other-molecules-in-venuss-atmosphere/; PI. D. Clements.

Cordiner, M. A., Villanueva, G. L., Wiesemeyer, H., Milam, S. N., de Pater, I., Moullet, A., et al. (2022). Phosphine in the Venusian Atmosphere: A Strict Upper Limit from SOFIA GREAT Observations. *Geophysical Research Letters*, e2022GL101055.

Currie, M. J., Berry, D. S., Jenness, T., Gibb, A. G., Bell, G. S., & Draper, P. W. (2014). Starlink software in 2013 (Vol. 485, p. 391).

Encrenaz, T., Greathouse, T. K., Marcq, E., Widemann, T., Bézard, B., Fouchet, T., et al. (2020). A stringent upper limit of the PH3 abundance at the cloud top of Venus. *Astronomy & Astrophysics*, *643*, L5.

Ferus, M., Cassone, G., Rimmer, P., Saija, F., Mráziková, K., Knížek, A., & Civiš, S. (2022). Abiotic chemical routes towards the phosphine synthesis in the atmosphere of Venus. In *European Planetary Science Congress* (pp. EPSC2022-198).

Garvin, J. B. et al. (2022) 'Revealing the Mysteries of Venus: The DAVINCI Mission', The Planetary Science Journal, 3(5), p. 117. doi: 10.3847/psj/ac63c2.

Glindemann, D., Bergmann, A., Stottmeister, U., & Gassmann, G. (1996). Phosphine in the lower terrestrial troposphere. Naturwissenschaften, 83(3), 131-133

Greaves, J. S., Richards, A. M. S., Bains, W., Rimmer, P. B., Sagawa, H., Clements, D. L., et al. (2021). Phosphine gas in the cloud decks of Venus. *Nature Astronomy*, *5*(7), 655–664.

Greaves, J. S., Rimmer, P. B., Richards, A., Petkowski, J. J., Bains, W., Ranjan, S., et al. (2022). Low levels of sulphur dioxide contamination of Venusian phosphine spectra. *Monthly Notices of the Royal Astronomical Society*, *514*(2), 2994–3001. https://doi.org/10.1093/mnras/stac1438

Lefèvre, M., Marcq, E., & Lefèvre, F. (2022). The impact of turbulent vertical mixing in the Venus clouds on chemical tracers. *Icarus*, *386*, 115148.

Mogul, R., Limaye, S. S., Lee, Y. J., & Pasillas, M. (2021). Potential for Phototrophy in Venus' Clouds. *Astrobiology*, *21*(10), 1237–1249. https://doi.org/10.1089/ast.2021.0032

Mogul, R., Limaye, S. S., Way, M. J., & Cordova, J. A. (2021). Venus' Mass Spectra Show Signs of Disequilibria in the Middle Clouds. *Geophysical Research Letters*, e2020GL091327.

Sousa-Silva, C., Seager, S., Ranjan, S., Petkowski, J. J., Zhan, Z., Hu, R., & Bains, W. (2020). Phosphine as a biosignature gas in exoplanet atmospheres. *Astrobiology*, *20*(2), 235–268.

Trompet, L., Robert, S., Mahieux, A., Schmidt, F., Erwin, J., & Vandaele, A. C. (2020). Phosphine in Venus' atmosphere: Detection attempts and upper limits above the cloud top assessed from the SOIR/VEx spectra. *Astronomy & Astrophysics*, *645*, L4.